\documentclass[preprint,showpacs,preprintnumbers,amsmath,amssymb]{revtex4}
\usepackage{graphicx}

\begin{document}

\title{Scanning tunneling spectroscopy characterization of the pseudogap and the $x=\frac{1}{8}$ anomaly in $La_{2-x}Sr_xCuO_4$
thin films}

\author{Ofer Yuli}
\affiliation{Racah Institute of Physics, The Hebrew University of
Jerusalem, Jerusalem 91904, Israel}

\author{Itay Asulin}
\affiliation{Racah Institute of Physics, The Hebrew University of
Jerusalem, Jerusalem 91904, Israel}

\author{Gad Koren}
\affiliation{Department of Physics, Technion - Israel Institute of
Technology, Haifa 32000, Israel}

\author{Oded Millo}
\email{milode@vms.huji.ac.il} \affiliation{Racah Institute of
Physics, The Hebrew University of Jerusalem, Jerusalem 91904,
Israel}

\begin{abstract}
Using scanning tunneling spectroscopy we examined the local
density of states of thin \textit{c}-axis $La_{2-x}Sr_xCuO_4$
films, over wide doping and temperature ranges. We found that the
pseudogap exists only at doping levels lower than optimal. For $x
= 0.12$, close to the 'anomalous' $x = \frac{1}{8}$ doping level, a
zero bias conductance peak was the dominant spectral feature,
instead of the excepted V-shaped (\textit{c}-axis tunneling) gap
structure. We have established that this surprising effect cannot
be explained by tunneling into (110) facets. Possible origins for
this unique behavior are discussed.

\end{abstract}


\maketitle
\section{Introduction}
\indent For a certain doping range, at temperatures higher than
the superconducting transition temperature $T_c$, hole doped
cuprate high temperature superconductors (HTSC) exhibit depletion
in the density of states (DOS) around the Fermi energy, $E_F$.
This 'soft gap', known as the pseudogap (PG) persists up to a
doping dependent temperature $T^* > T_c$. The PG phenomenon is not
yet well understood and it is even still unclear whether its
origin is related to superconductivity or not. Experimentally, the
PG has been studied extensively for different HTSCs, using various
methods. \cite{Timusk} Several experiments show that the PG exists
only in the underdoped regime and that there, for each doping
level, it evolves smoothly into the superconducting (SC) gap at
the corresponding $T_c$. This suggests that both phases share a
similar energy scale. In contrast, quite a few studies reveal
deviations from the above 'universal' behavior. Angle resolved
photo emission spectroscopy (ARPES) investigations
\cite{Ino1,Sato} imply that there may be two PGs of different
origins in $La_{2-x}Sr_xCuO_4$ (LSCO), with typical gap values of
$\sim$ 30 and $\sim$ 100 meV. Scanning tunneling microscopy and
spectroscopy (STM and STS) studies of Bi-based cuprates show that
the PG extends into the overdoped regime, \cite{Renner,Kugler}
while no PG has been measured for optimally- and overdoped
$YBa_2Cu_3O_{7-\delta}$ (YBCO). \cite{Aprile} These contradicting
data raise the question of whether the observed results are truly
universal attributes of the PG in the HTSC cuprates.\\
\indent STS enables precise measurements of the single particle
excitation spectra, with high energy and spatial resolution,
making it a powerful tool for studying the energy scales of both
the SC gap and the PG.  Atomically-resolved STS measurements of
the DOS in $Bi_2Sr_2CaCu_2O_{8-\delta}$ (BSCCO) revealed large
spatial fluctuations of the quasi-particle DOS
\cite{Davis,McElroy,Yazdani,Kapitulnik} and provided the first
tunneling spectroscopy data of the PG doping and temperature
dependencies. \cite{Renner} In fact, most of the STS
investigations of the cuprates has been performed on BSCCO due to
the relative ease of achieving a clean surface by cleavage under
vacuum. In contrast, very few STM studies have been carried out on
LSCO due to problems of surface degradation. The same problem also
hindered STM measurements of heavily underdoped YBCO.\\
\indent The relatively low $T_c$ and ease of sample preparation
over a wide doping range make LSCO an ideal candidate for studying
the PG. This is illustrated by the abundant experimental data
available, acquired using various techniques such as transport
measurements, \cite{Curra} point contact spectroscopy
\cite{Gonnelli}, tunnel junctions, \cite{Dagan} infra red
spectroscopy, \cite{Wang} and ARPES. \cite{Damascelli} It is
therefore unfortunate that STS measurements on LSCO are relatively
scarce, and were performed mainly on optimally doped samples.
Obviously, it is very important to expand the STS measurements in
LSCO also to the under- and overdoped regimes of the phase
diagram, to obtain a comprehensive picture of the doping
dependence of the DOS and elucidate the relation between the SC
and PG states. This will hopefully be an important step towards
resolving the mechanism underlying high $T_c$ superconductivity.\\
\indent Another unresolved property of LSCO and related cuprates
(e.g. the La-214 family) is the $x = \frac{1}{8}$ anomaly, where a
suppression of the superconducting transition temperature is
observed. This is manifested in $La_{2-x}Ba_xCuO_4$ \cite{Mood}
where $T_c$ drops to nearly zero at $x = \frac{1}{8}$, and in
LSCO, in a more subtle way, where a plateau develops in the $T_c$
versus doping curve, $T_c$(x). \cite{Matsuzaki,Sato2} In addition,
magnetization and penetration depth results have recently been
reported to behave anomalously at $x = \frac{1}{8}$. \cite{Png}
The above phenomena have been attributed by various authors to a
stripe phase known to exist in these compounds. \cite{Orgad} The
connection with stripes emerges from the fact that at $x =
\frac{1}{8}$ the stripe spacing is commensurate with the
underlying crystal structure. We note in passing, that anomalies
are expected also for other $x$ values at which commensurability
conditions prevail, often referred to as magic numbers. \cite{magic}\\
\indent Following the above considerations, we performed a
systematic STS study of LSCO over a wide range of doping and
temperature. To the best of our knowledge, this is the first
report on STS measurements of the temperature evolution of the DOS
in underdoped and overdoped LSCO, and for the $x = \frac{1}{8}$
doping level in particular. Our main findings are that the PG in
LSCO pertains only to the underdoped regime, and is absent for
doping levels higher than optimal. Additionally, for
\textit{c}-axis films with $x = 0.12$, a zero bias conductance
peak (ZBCP) was observed on large smooth \textit{c}-axis areas of
the sample surface, replacing the excepted V-shaped
(\textit{c}-axis tunneling) gap. This is a surprising result,
since the ZBCP is known to be the dominant spectral feature on
(110) surfaces due to the formation of Andreev bound states, while
in \textit{c}-axis tunneling spectra no such ZBCP is expected.
\section{Experiment}
\indent 90 nm thick LSCO films were epitaxially grown on (100)
$SrTiO_3$ wafers by laser ablation deposition, with
\textit{c}-axis orientation normal to the substrate. About three
films for each of the following doping levels were measured: $x =
0.08, 0.10, 0.12$ (underdoped); $x = 0.15$ (optimally doped) and
$x = 0.18$ (overdoped). The LSCO films consisted of
\textit{c}-axis crystallites having well defined facets, 50 - 100
nm long, as shown in Fig. 1(a). As will be detailed below, most of
the exposed side facets are (100) oriented, but some exposed side
facets are of the (110) surface. The samples were transferred from
the growth chamber in a dry atmosphere and introduced into our
cryogenic STM after being exposed to ambient air for less than 5
min. About six tunneling spectra (dI/dV vs. V characteristics)
were acquired at each position to assure data reproducibility. We
also checked the dependence of the tunneling spectra on the
voltage and current settings (i.e., the tip-sample distance, or
the tunneling resistance values, in the range of 100 M$\Omega$ to
1 G$\Omega$) and found no effect on the measured gap features,
disproving the possibility that these gaps are even partially
related to single electron charging effects. \cite{Sadeh}  The
temperature dependent resistance [R(T)] measurements were
performed using the standard 4-probe technique, where special care
was taken to stabilize the temperature before each resistance
measurement and to avoid sample heating.
\begin{figure}
\includegraphics[width=8cm]{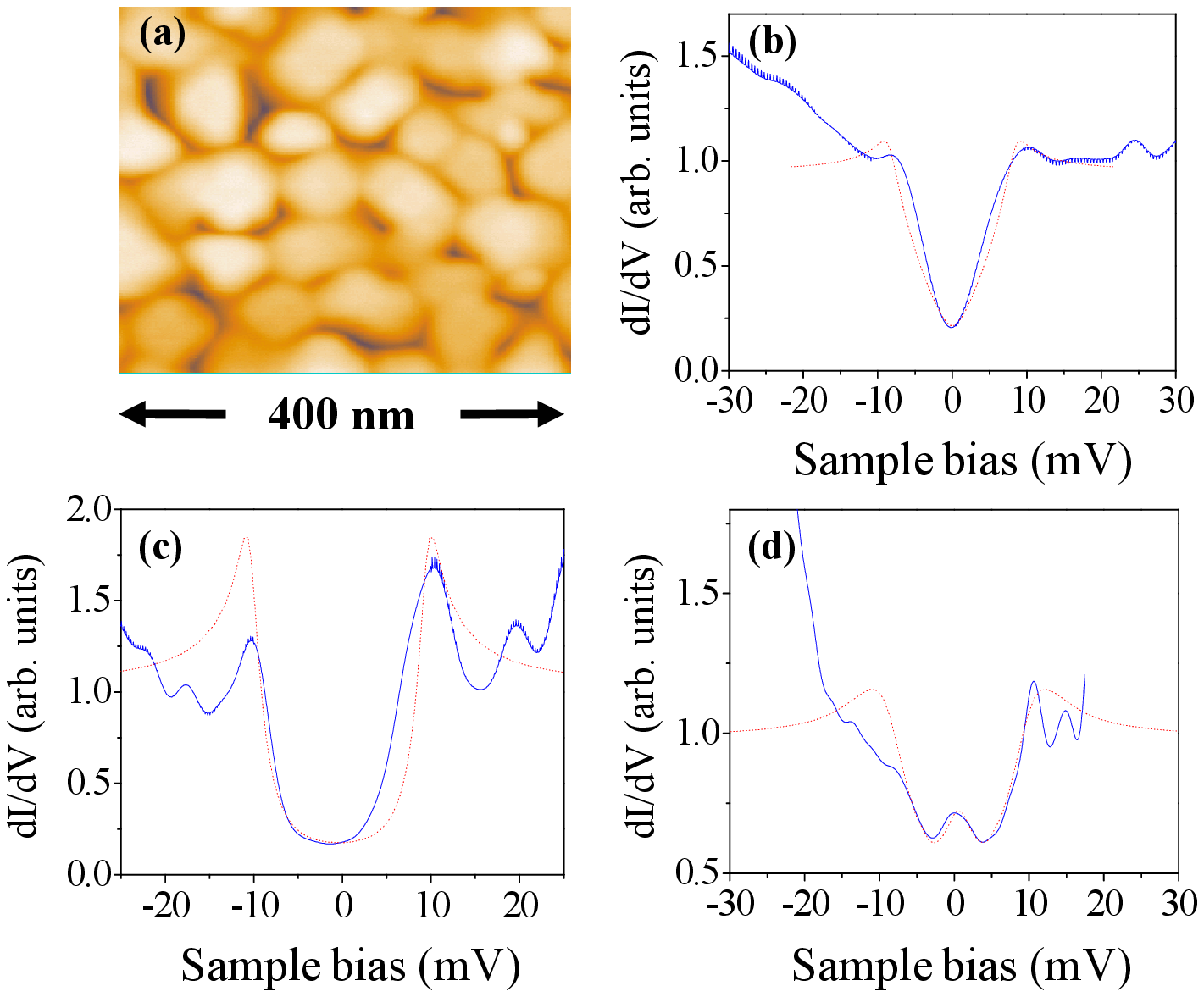}
\caption{(a) 400x400 $nm^2$ topography image of $x = 0.08$ LSCO
film. (b-d) Tunneling spectra at 4.2 K (blue curves) taken on a
single crystallite. The spectrum in (b) was measured a top of the
crystallite and shows a typical \textit{c}-axis V-shaped gap
structure. The spectra in (c) and (d) were acquired on two
different facets of the crystallite. The one in (c) portrays a
U-shaped gap, typical for tunneling into a \textit{smooth} (100)
facet, and that in (d) exhibits a ZBCP, signifying that the
tunneling took place out of the anti-nodal direction.  The dotted
red lines are fits to the theory of tunneling into a
\textit{d}-wave SC. }\label{fig1}
\end{figure}
\section{RESULTS AND DISCUSSION}
\subsection{Evolution of the gap structure with doping and temperature}
\indent For all the doping levels studied, except for $x = 0.12$
(see below), the tunneling spectra measured on top of the LSCO
crystallites exhibited V-shaped gap structures typical of
\textit{c}-axis tunneling, further confirming the assigned
crystallographic orientation found by X-ray diffraction. A typical
spectrum is presented in Fig. 1(b) along with a fit to the model
for tunneling into a \textit{d}-wave superconductor. \cite{TK} In
many cases, the spectra acquired on the crystallite side facets
portrayed a U-shaped gap structure [see Fig. 1(c)], signifying
tunneling into \textit{smooth} (100) facets, as discussed in Ref.
24.  Finally, as depicted by Fig. 1(d), along some of the side
facets, we have measured ZBCPs (usually inside a gap-like
structure), suggesting that the corresponding facet exposes the
(110) surface, or at least that tunneling takes place in the
\textit{a-b} ($CuO_2$) plane in a direction different from [100].
Indeed, the dotted red line in Fig. 1(d) was calculated using the
aforementioned model \cite{TK} assuming tunneling to the
\textit{a-b} plane at an angle of $\frac{\pi}{3}$ with respect to
the [100] axis. A similar correlation between the tunneling
spectra and the surface nano-morphology was observed in our
previous investigations of YBCO films. \cite{Sharoni1}  The ZBCP,
a unique spectral feature of \textit{d}-wave SCs, \cite{Hu} will
be further discussed below. Interestingly, pronounced above-gap
structures appear in the tunneling spectra that were aquired on
the side facets of the \textit{c}-axis crystallites [spectra (c) and
(d)]. Such features occasionally appear, in a weaker manner, also in
the \textit{c}-axis (on crystallite) tunneling data. These
above-gap features may be related to the phonon structure, which
may be more efficiently
detected when tunneling into the \textit{a-b} plane. \cite{phonon}\\
\begin{figure}
\includegraphics[width=8cm]{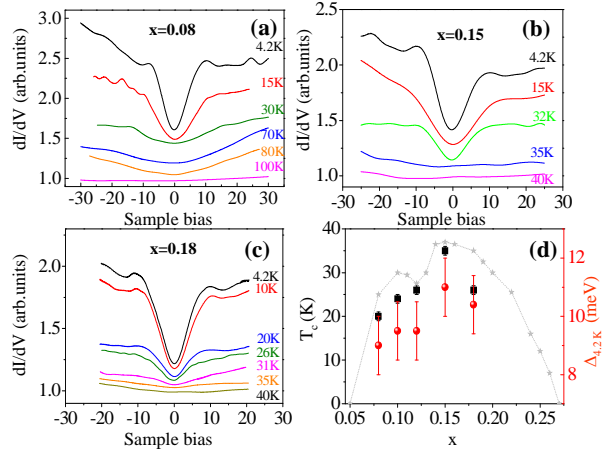}
\caption{(a-c) Tunneling dI/dV vs. V curves (shifted vertically
for clarity) at various temperatures and for three different
doping levels: (a) $x = 0.08$ (underdoped, $T_c \approx$ 20 K);
(b) $x = 0.15$ (optimally doped, $T_c \approx$ 35 K); (c) $x =
0.18$ (overdoped, $T_c \approx$ 29 K). (d) The 4.2 K gap (red
circles, right axis) and  $T_c$ (black squares, left axis) as a
function of doping. For comparison we plot $T_c$(x) measured on
LSCO single crystals from T. Matsuzaki, Phys. Chem. Sol.
\textbf{62}, 29 (2001).}\label{fig2}
\end{figure}
\indent In fig 2 we present differential conductance spectra
acquired at selected temperatures on three samples: a heavily
underdoped $x = 0.08$ (a), an optimally doped $x = 0.15$ (b) and
an overdoped $x = 0.18$ (c), samples. The gap value, $\Delta$,
was extracted from the data by taking half the width between the
coherence peaks. At temperatures higher than 4.2 K, where the
coherence peaks were smeared, the maximum change of the slope was
taken as the position of the gap edge. This method, which yields
some uncertainty in the determination of the gap values, was
applied before for tunneling spectra obtained on LSCO, since in
many cases the coherence peaks were undetectable. \cite{Kato} Fig.
2(d) summarizes the $\Delta_{4.2 K}$ values (red circles, right
axis) and the measured $T_c$ values (black rectangles, left axis)
for all the doping levels measured. The error bars associated with
the gap data represent the range of spatial fluctuations in the
$\Delta_{4.2 K}$ values, whereas those corresponding to $T_c$
reflect the width of the transition. For reference we have added
the $T_c(x)$ values measured on single crystals (asterisks and
dotted line) from Ref. 17. The systematic lower $T_c$ values
measured on our thin films, as compared to those of the single
crystals, is probably due to strain forces generated by the
mismatch between the lattice constants of the $SrTiO_3$ substrate
and the LSCO film. \cite{Bozovic,Sato2} The evolution of $\Delta$
with doping appears to follow the SC dome structure. This appears
to be a reasonable result, if the gap magnitude is related to the
pair potential of the Cooper pairs that, in turn, is associated
with $T_c$. It should be emphasized, however, that there are
reports of STS data showing different behaviors. \cite{Renner,NCY}
We note in passing that our observation for the doping dependence
of $\Delta_{4.2 K}$ is similar to that of the peak in the Raman
spectra associated with nodal-quasiparticles excitations, reported
in Ref 31. The origin of this correspondence is not yet clear to
us. Moreover, we do not find evidence for the two energy scales
discussed in the latter paper, maybe because \textit{c}-axis
tunneling spectroscopy
probes \textit{simultaneously} the anti-nodal and nodal excitations.\\
\begin{figure}
\includegraphics[width=8cm]{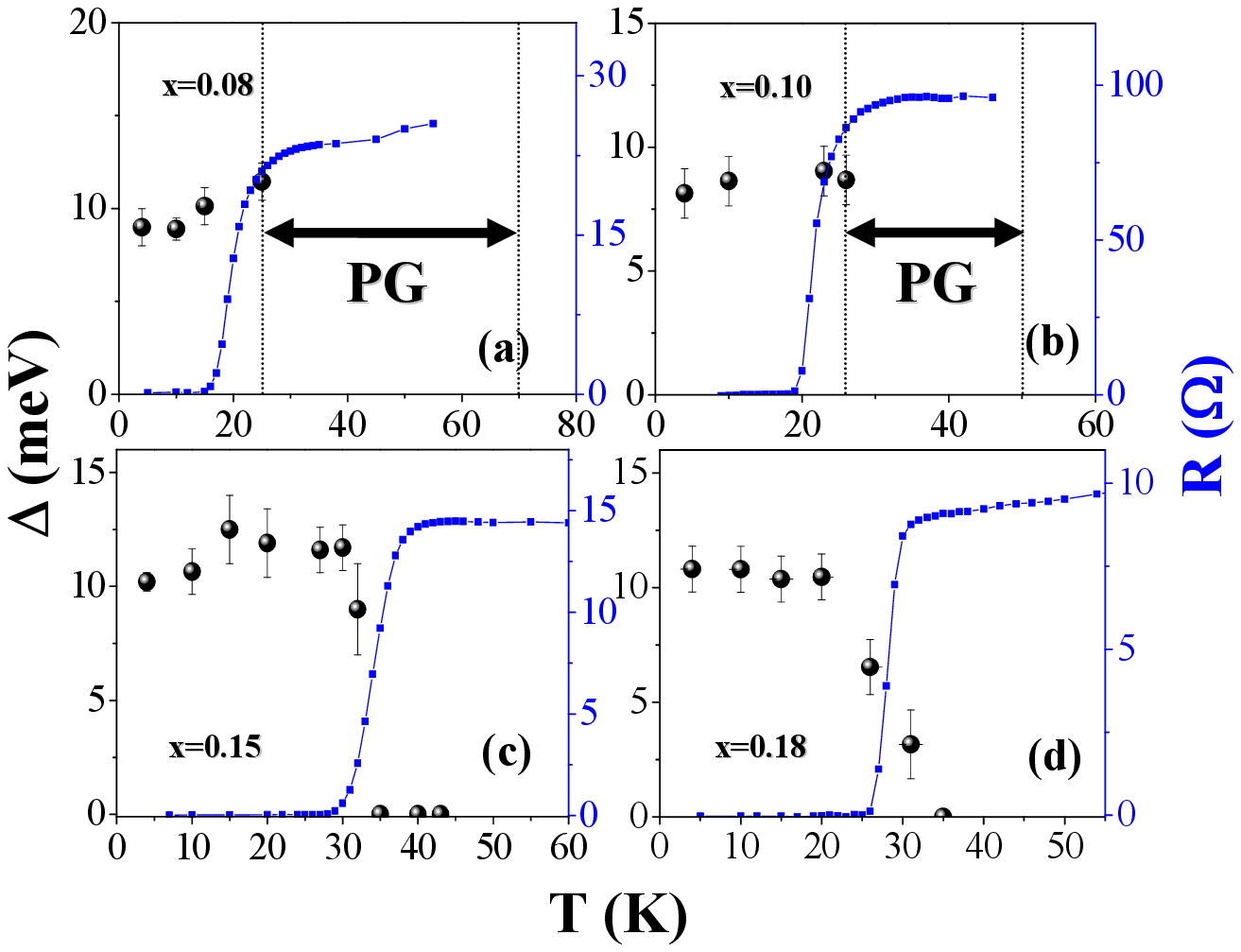}
\caption{Summary of the SC gap width evolution with
temperature for underdoped samples, $x = 0.08$ (a) and $x = 0.10$
(b), optimally doped, $x = 0.15$ (c) and overdoped, $x = 0.18$
(d) samples. The gaps (black circles, left axes) are plotted along
with the R(T) data (blue squares and lines, right axes) for each
doping level. The double sided arrow in (a) and (b) marks the 
temperature range where the PG was clearly detectable as a depression 
in the tunneling spectra. At the right hand side of the double sided 
arrow ($T^*$) the zero-bias conductance increases toward the normal 
conductance and consequently the gap structure vanishes. These data 
show that the PG exists only for underdoped samples.}\label{fig3}
\end{figure}
\indent Fig 3 summarizes the evolution with temperature of the SC
gap width (below $T_c$) for different doping levels. The gap
values (black circles, left axes) were extracted from the
differential conductance curves, as explained above. The error
bars represent standard deviations of the spatial distribution of
the gap magnitude and the error in determining the gap value when
the coherence peaks are absent, as discussed above. For each
doping level we present the corresponding R(T) curve that was
measured right after the STM measurements. As seen in Fig. 2(a),
the SC gap of the underdoped $x = 0.08$ films does not close up
at $T_c$, as a BCS gap would. Instead, the gap evolved smoothly
into a PG, manifested by a depletion in the density of states
around $E_F$. The PG that was noticeable up to $T^{*} \sim$ 70 K, well
above the onset temperature of superconductivity, $T_c^{onset}
\sim$ 30 K. The smearing of the gap edges at $T > T_c$ created
large errors in the determination of the (pseudo)gap width,
therefore we do not plot the gap values. Nevertheless, the
temperature range where the PG was clearly detectable is presented
in Fig. 3(a) as a double sided arrow, starting at $T_c$ and ending
at the temperature, $T^{*}$, where the spectra turned gapless (or
Ohmic). Qualitatively, a similar behavior was observed for the
(still underdoped) $x = 0.10$ samples, with $T^{*} \sim$ 50 K [Fig. 3(b)].\\
\indent The results of the optimally doped and overdoped films
are shown in Figs. 3(c) and 3(d).  As in the case of the
underdoped samples, the SC-gap width remains almost constant
nearly up to $T_c$. However, in contrast with the underdoped
films, the gap in the DOS closes up and vanishes near $T_c$, while
no sign of the PG is observed. The behavior of the overdoped ($x
= 0.18$) differs even more profoundly from that of the underdoped
samples. As can be seen in Figs. 2(c) and 3(d), the width of the
SC gap (as well as its depth) reduced significantly already at
$\sim \frac{2}{3}T_c^{onset}$, and completely disappears at
$T_c^{onset} \sim$ 31 K. This behavior resembles the temperature
evolution of a BCS-like gap. \cite{Tinkham} Our data may thus
portray a crossover from a non-BCS to a BCS-like SC gap behavior
as the doping increases from below to above optimal doping, as
suggested previously. \cite{Deutscher,Uemura}\\
\indent We conclude that our STS results corroborate earlier
findings on LSCO (using other experimental methods) as well as on
other cuprates, showing that the PG is an intrinsic property of
the underdoped regime. However, both the doping dependence of the
gap magnitude and the restriction of the PG to the underdoped
regime exclusively, are in contrast to findings by Renner et al.
for BSCCO. \cite{Renner} These authors found clear evidence for a
PG in overdoped BSCCO samples (while we do not) and reported that
$\Delta$ monotonically decreased as the doping level was increased
(whereas we found that $\Delta$ follows the SC dome, see Fig. 2).
These discrepancies lead one to question the universality of the
SC gap and PG attributes among the family of the HTSC cuprates. We
note that the cuprates differ from one another in various ways,
e.g., in the number of $CuO_2$ planes within the unit cell, the
$T_c$ values and the doping methods, thus universality should not
necessarily prevail.
\subsection{Anomalous behavior at x = 0.12}
\indent When examining \textit{c}-axis LSCO films with $x = 0.12$
 $(\approx \frac{1}{8})$ we encountered a surprising effect. The
local tunneling spectra over large areas of the sample, including
those measured \textit{on top} of \textit{c}-axis crystallites,
exhibited ZBCPs instead of the expected \textit{c}-axis V-shaped
tunneling gap. In Fig. 4 we present a typical line scan taken on
top of a \textit{c}-axis crystallite (shown in the inset) in one
of our $x = 0.12$ films. All the spectra indeed exhibit clear ZBCP
structures.  This behavior was observed for all three $x = 0.12$
samples we measured, on different locations on each film separated
by a few millimeters from one another and for various directions
of the line-scan. Areas showing the expected (ZBCP-free) gap
structure were also found, but they comprised less than 15\% of
the scanned sample area [the corresponding data were used in the
analysis shown in Fig. 2(d)]. We would like to emphasize that the
unexpected appearance of the ZBCP was exclusively found in the $x
= 0.12$ doping level, whereas for other doping levels only
V-shaped gaps were observed on top of the crystallites (see Fig.
1). In the latter $(x \neq 0.12)$ films, the ZBCPs were detected
only along some of the crystallite facets, as discussed above.\\
\begin{figure}
\includegraphics[width=8cm]{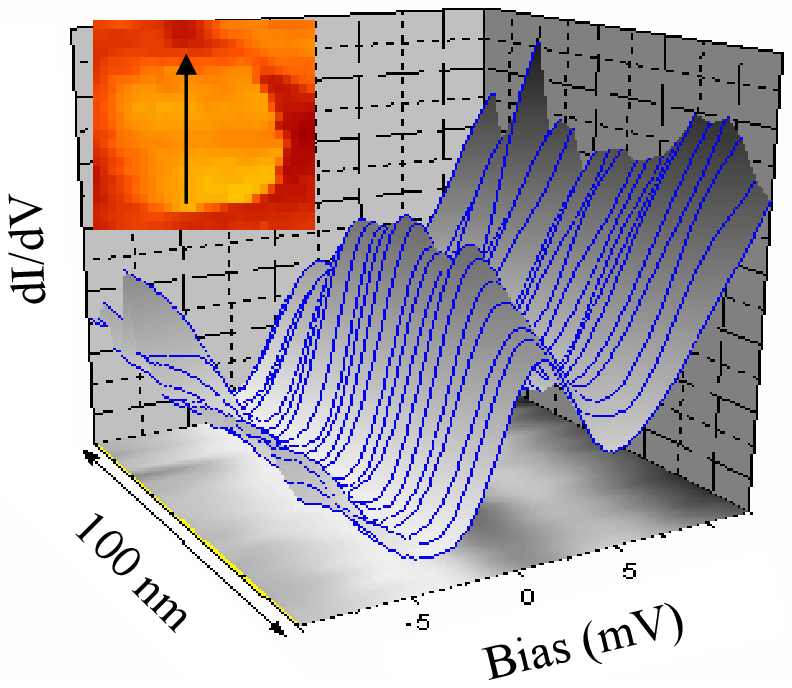}
\caption{Tunneling spectra measured at 4.2 K on a \textit{c}-axis
$x = 0.12$ LSCO film. The spectra were taken sequentially at
equidistant steps ($\sim$ 4 nm) along the 100 nm long line running
on top of a \textit{c}-axis crystallite, as shown in the inset.
Surprisingly, the spectra exhibit a ZBCP instead of a pure
V-shaped gap structure. }\label{fig4}
\end{figure}
\begin{figure}
\includegraphics[width=8cm]{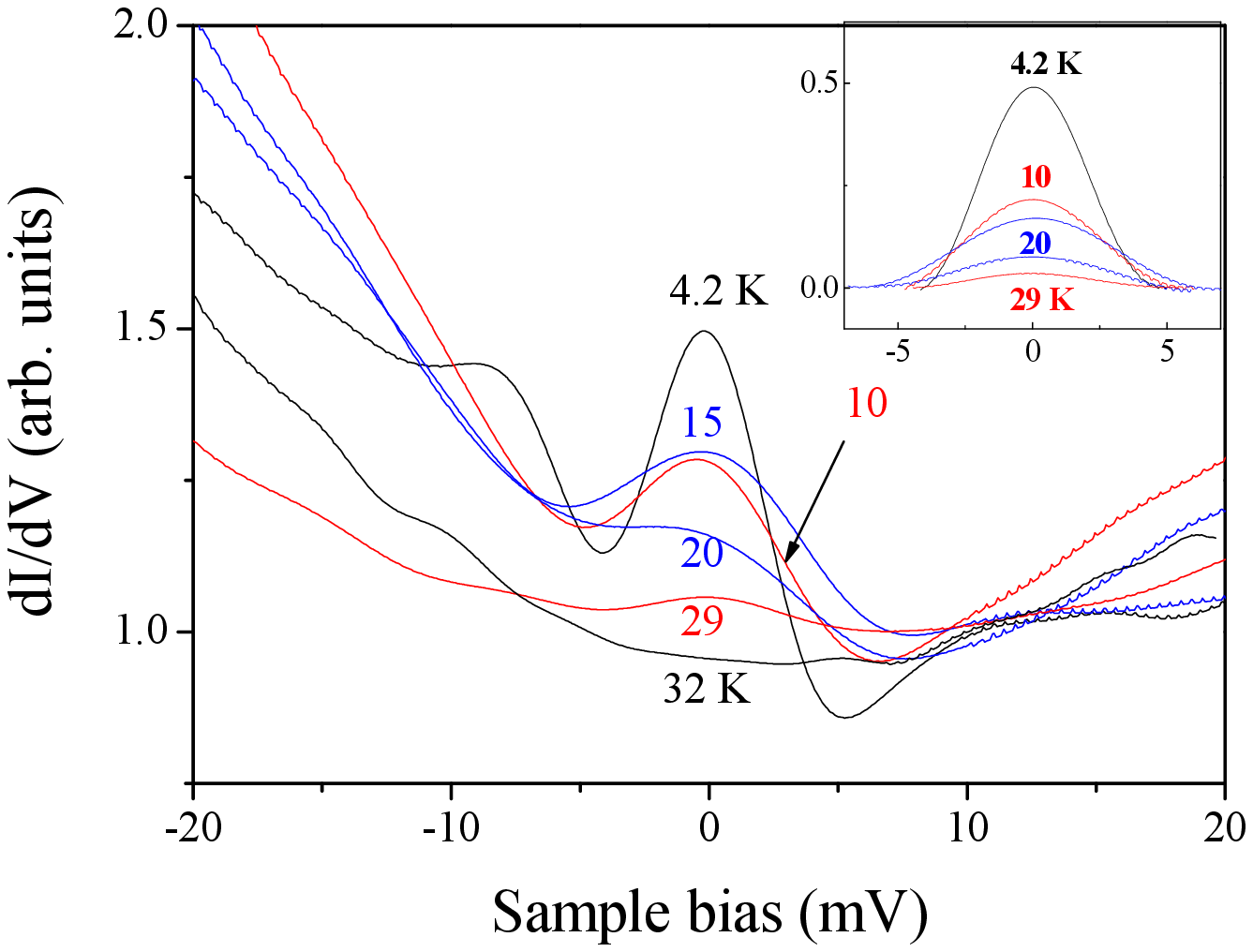}
\caption{Tunneling spectra curves at various temperatures measured
on an $x = 0.12$ LSCO film.  Inset: The ZBCPs after background
subtraction.}\label{fig5}
\end{figure}
\indent The ZBCP spectral feature is one of the hallmarks of
\textit{d}-wave superconductivity. \cite{Hu,TK} It originates from
the formation of zero-energy (at $E_F$) surface bound states,
known as Andreev bound states (ABS). The ABS appear on the nodal
surfaces of \textit{d}-wave superconductors [(110) in LSCO and
YBCO] due to constructive interference of multiple Andreev
reflected quasiparticles that experience a sign change of the
order parameter in consecutive reflections. \cite{Hu,TK} The ZBCP
is \textit{not} expected to be observed in spatially resolved STS
data acquired on top of \textit{c}-axis crystallites, but can be
detected on side facets of these crystallites, as demonstrated
above and even more clearly on pure (110) facets. \cite{Sharoni1}\\
\indent We believe that the anomalous abundance of ZBCPs  reported
here for the $x = 0.12$ films cannot be associated with faceting
and local tunneling along the nodal [110] direction (in spite of
the fact that they look very similar to the nodal, ABS-related,
ZBCPs). As noted above, line scans continuously showing ZBCPs were
measured along different directions (that did not necessarily run
along any major crystallographic direction) on a single
crystallite. In addition, no ZBCP were found on-top of
crystallites in the films of other ($x \neq 0.12$) doping levels,
indicating that the presence of (110) nano-facets  is negligible
in our films. These findings practically rule out the possibility
that the ZBCPs are associated with (110) facets. Moreover, X-ray
diffraction measurements on our samples revealed highly oriented
\textit{c}-axis films with sharp (00n) peaks and no other orientations.\\
\indent Next we address the temperature evolution of the ZBCP in
the $x = 0.12$ films. As demonstrated in Fig. 5, the magnitude of
the ZBCP gradually decreases as the temperature is raised and
vanishes at about 30 K, very close to (and slightly above) the SC
transition temperature. In order to compare the 'anomalous' ZPCPs
with the 'conventional' ones associated with the nodal ABS, we
have performed temperature dependent STS measurements on $x =
0.12$ LSCO films grown intentionally on (110) STO wafers. X-ray
diffraction measurements showed that in addition to the (110) LSCO
orientation, a large amount of (103) orientation is also present
in these films. This however posed no problem since, as previously
shown, ZBCPs due to ABS exist also for this orientation.
\cite{Greene,Lesueur} Fig. 6 displays a comparison between the
temperature evolution of the ZBCP magnitude measured on (001) and
nominal (110) films. Each data point in this figure represents the
integrated area under the corresponding ZBCP after subtraction of
the background conductance as shown in the inset of Fig. 5.
Assuming that the ZBCP observed on the \textit{c}-axis film is
also due to some type of zero-energy surface bound states, the
data in Fig. 6 reflect the number of bound states at each
temperature. \cite{Asulin} Both sets of data show that the ZBCP
magnitude decreases monotonically as the temperature is increased.
There is, however, a marked difference between the two samples,
namely, the two types of ZBCPs.  While the ZBCPs on the (001)
surface were still detectable at around $T_c \approx 27 K$, the
'conventional' ZBCPs vanished already at a much lower
temperature, $\sim$ 18 K, well below $T_c$.\\
\begin{figure}
\includegraphics[width=8cm]{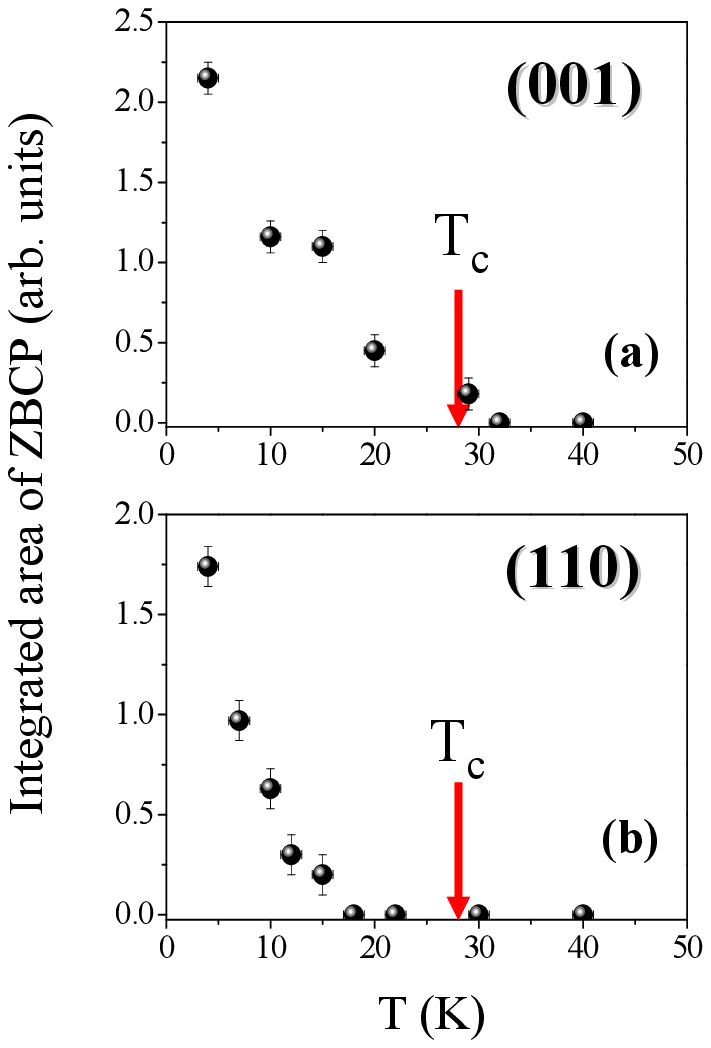}
\caption{The temperature dependence of the spectral weight the
ZBCPs measured on the (001) (a) and (110) (b) films. The (001)
ZBCP vanished at around $T_c$,  whereas that measured on the (110)
film vanishes at a lower temperature, as commonly found for ZBCPs
associated with the 'conventional' nodal ABS.}\label{fig6}
\end{figure}
\indent Previous tunneling spectroscopy investigations of the
nodal surfaces of LSCO and YBCO showed that the ABS usually lose
all spectral weight at about $0.5 - 0.6 T_c$. \cite{Dagan,Greene2}
In particular for the (110) surface of an $x = 0.12$ LSCO single
crystal, Dagan et al. have shown, using point contact
spectroscopy, that the ZBCP disappeared at $\sim$ 15 K, while
$T_c$ was $\sim$ 31 K. The qualitative agreement of this finding
with our results for the nominal (110) surface [Fig. 6(b)] on one
hand, and the different behavior we found for the 'anomalous'
(001)-surface ZBCPs [Fig. 6(a)] on the other hand, further support
our claim that the $x = 0.12$ anomaly is not a faceting effect.
Nevertheless, the fact that the 'anomalous' ZBCPs have a similar
shape to that of the nodal ones and that they also quench with
increasing temperature suggests that they also manifest surface
bound states resulting from multiple Andreev reflections. The
exact mechanism (or multiple Andreev reflection process) that
underlines the formation of these bound states is probably
different from that associated with the nodal ABS \cite{Hu} and
has not been predicted or discussed in the literature as of yet.\\
\indent Bobkova et al. predicted theoretically that zero-energy
bound states should exist at the interface between the
\textit{a-b} plane of a \textit{d}-wave SC and a charge density
wave (CDW) material, provided this junction has high enough
transparency. \cite{Bobkova} These bound states represent a
combined effect of Andreev reflections from the superconducting
side, unconventional Q-reflections from the CDW solid, and
standard specular reflections. In a Q-reflection from the
interface with the CDW solid, a sub CDW-gap quasiparticle changes
its momentum by the wavevector Q of the CDW pattern. While this
model is not directly related to our experimental configuration,
it may indicate a possible direction at which a relevant model
could be established, namely, the involvement of static stripes.
Neutron scattering measurements have shown the existence of static
stripes commensurate with the underlying crystal structure in LSCO
for $x = \frac{1}{8}$. \cite{Orgad} These stripes exhibit charge
modulation that can be related to a CDW order parameter, which
coexists with a \textit{d}-wave order parameter. A serious
theoretical investigation is thus needed on the possible role of
stripes in the appearance of the ZBCP
anomaly that we observed in the $x = \frac{1}{8}$ LSCO films.\\
\section{summary}
\indent In conclusion, we have shown that the PG in epitaxial thin
films of LSCO exists only in the underdoped regime where above
$T_c$, the gap became more and more shallow and eventually
vanished at a doping dependent temperature $T^{*} > T_{c}$ . At
4.2 K the SC gap magnitude was found to qualitatively follow the
SC dome structure versus doping, as expected for a gap magnitude
proportional to the pairing potential. The disagreement of our
results with those obtained on BSCCO suggests that the SC gap and
PG properties may not be universal among the HTSC cuprates. We
have also revealed a new anomaly associated with the special $x =
\frac{1}{8}$ doping level of LSCO.  The \textit{c}-axis tunneling
spectra exhibited a predominant ZBCP instead of the expected
V-shaped gap, manifesting the creation of a new type of zero
energy bound states. Ruling out the possibility that this is a
facet effect, we conjecture that this anomalous spectral feature
is associated with static stripes. Further theoretical modeling
and experimental data using different methods are needed in order
to resolve the puzzle associated with
this intriguing effect.\\
\section{Acknowledgments}
\indent The authors are grateful to G. Deutscher, D. Orgad, and S.
Baruch, for helpful and stimulating discussions. This research was
supported by the Israel Science Foundation, Center of Excellence
program (grant \# 1565/04).

\end{document}